\documentclass{aa}
\usepackage{graphicx}
\usepackage[normalem]{ulem}
\begin{document}

\newcommand\BB{{\overline{B}}}
\newcommand\HH{{B}}
\newcommand\bb{{b}}
\newcommand{\cs}{c_{\rm s}}
\newcommand\deriv[2]{\displaystyle\frac{\partial #1}{\partial #2} }
\newcommand{\HI}{{\rm H\,\scriptstyle I}}
\newcommand\N{{N}}
\newcommand\n{{n}}
\newcommand{\ncr}{n_{\rm cr}}
\newcommand{\nel}{n_{\rm e}}
\newcommand{\ntot}{n_{\rm t}}
\newcommand\Pgas{{{\cal P}_{\rm gas}}}
\newcommand{\RM}{{\rm RM}}
\newcommand\sfrac[2]{{\textstyle{\frac{#1}{#2}}}}

%
%
\newcommand{\cm}{\,{\rm cm}}
\newcommand{\cmcube}{\,{\rm cm^{-3}}}
\newcommand{\dyn}{\,{\rm dyn}}
\newcommand{\erg}{\,{\rm erg}}
\newcommand{\Jy}{\,{\rm Jy}}
\newcommand{\Jyb}{\,{\rm Jy/beam}}
\newcommand{\kms}{\,{\rm km\,s^{-1}}}
\newcommand{\mJy}{\,{\rm mJy}}
\newcommand{\mJyb}{\,{\rm mJy/beam}}
\newcommand{\K}{\,{\rm K}}
\newcommand{\kpc}{\,{\rm kpc}}
\newcommand{\Mpc}{\,{\rm Mpc}}
\newcommand{\mG}{\,{\rm mG}}
\newcommand{\mkG}{\,\mu{\rm G}}
\newcommand{\MHz}{\, {\rm MHz}}
\newcommand{\Msol}{\,{\rm M_\odot}}
\newcommand{\p}{\,{\rm pc}}
\newcommand{\radm}{\,{\rm rad\,m^{-2}}}
\newcommand{\s}{\,{\rm s}}
\newcommand{\yr}{\,{\rm yr}}

\title{Systematic bias in interstellar magnetic field estimates}

\author{
Rainer~Beck \inst{1}
\and
Anvar Shukurov \inst{1,2}
\and
Dmitry Sokoloff\, \inst{3}
\and
Richard Wielebinski \inst{1}
}

\offprints{R.~Beck}

\authorrunning{R.~Beck et al.}
\titlerunning{Systematic bias in interstellar magnetic field estimates}

\institute{
Max-Planck-Institut f\"ur Radioastronomie, Auf dem H\"ugel 69, D-53121
Bonn,
Germany
\and
School of Mathematics and Statistics, University of Newcastle,
Newcastle upon Tyne, NE1 7RU, UK
\and
Department of Physics, Moscow State University, 119992 Moscow, Russia
}

\date{Received 6 March 2003 / Accepted 1 July 2003}

\abstract{
Faraday rotation of the polarization plane in magnetized thermal plasma
provides one of the most efficient methods to deduce regular magnetic fields
from radio astronomical observations. Since the Faraday rotation measure $\RM$
is proportional to an integral, along the line of sight, of magnetic field
weighted with thermal electron density, $\RM$ is believed to yield the regular
magnetic field averaged over large volume.  Here we show that this is not the
case in a turbulent medium where fluctuations in magnetic field and electron
density are not statistically independent, and so contribute to $\RM$. For
example, in the case of pressure equilibrium, magnetic field can be
anticorrelated with plasma density to produce a negative contribution. As a
result, the strength of the regular magnetic field obtained from $\RM$ can be
{\em underestimated\/} if the fluctuations in electron density and magnetic
field are neglected. The anticorrelation also reduces the standard deviation
of $\RM$.  We further discuss the effect of the positive correlations where
the standard treatment of $\RM$ leads to an {\em overestimated\/} magnetic field.
Because of the anisotropy of the turbulent magnetic field,
the regular magnetic fields strength, obtained from synchrotron emission using
standard formulae, can be {\em overestimated}.
A positive correlation between cosmic-ray number density and magnetic
field leads to an overestimate of the strengths of the regular and total fields.
These effects can explain the
difference between the strengths of the regular Galactic magnetic field as
indicated by $\RM$ and synchrotron emissivity data and reconcile the magnetic
field strength in the Solar vicinity with typical strength of regular magnetic
fields in external galaxies.

}

\maketitle

\section{Introduction}
Estimates of magnetic field strength in the diffuse interstellar medium (ISM)
of the Milky Way and other galaxies are most efficiently obtained from the
intensity and Faraday rotation of synchrotron emission. Other methods are only
sensitive to relatively strong magnetic fields that occur in dense clouds
(Zeeman splitting) or are difficult to quantify (optical polarization of star
light by dust grains). The total $I$ and polarized $P$ synchrotron intensities
and the Faraday rotation measure $\RM$ are integrals over the path length $L$,
so they provide a measure of the average magnetic field in the emitting or
magneto-active volume:
\begin{eqnarray}
I&=&K\int_L\ncr\HH_\perp^2\,ds\;,\nonumber\\
P&=&K\int_L\ncr\BB_\perp^2\,ds\;,	\label{ints}\\
\RM&=&K_1\int_L\nel\HH_\parallel\,ds\;\nonumber,
\end{eqnarray}
where $\ncr$ and $\nel$ are number densities of relativistic and thermal
electrons,
$\vec{\HH}$ is the total magnetic field comprising a regular
$\vec{\BB}$ and random $\vec{\bb}$ parts,
$\vec{\HH}=\vec{\BB}+\vec{\bb}$ with $\langle\vec{\HH}\rangle=\vec{\BB},\
\langle\vec{\bb}\rangle=0$ and $\langle{\HH^2}\rangle=\BB^2+\langle\bb^2\rangle$,
angular brackets denote averaging, subscripts $\perp$ and $\parallel$
refer to magnetic field components perpendicular and parallel to the line of
sight, and $K$ and
$K_1=0.81\radm\cm^3\mkG^{-1}\p^{-1}$ are certain dimensional constants.
The degree of polarization $p$ is related to the ratio
$\langle\bb^2\rangle/\BB^2$,
\begin{equation}                \label{polar}
p = \frac{P}{I} \approx
p_0 \frac{\BB_\perp^2}{\HH_\perp^2}
=p_0\frac{\BB_\perp^2}{\BB_\perp^2+\sfrac23\langle\bb^2\rangle}\;,
\end{equation}
where the random field $\vec{\bb}$ has been assumed to be isotropic, $\ncr$ is
assumed to be a constant, and $p_0\approx0.75$ weakly depends on the spectral
index of the emission (Burn \cite{b66}; Sokoloff et al.\ \cite{S98}). This is
an approximate relation. In particular, it does not allow for any anisotropy
of the random magnetic field (see Sect.~\ref{AMF}), for equipartition between
magnetic fields and cosmic rays (see Sect.~\ref{LE}) and for depolarization
effects; some generalizations are discussed by Sokoloff et al.\ (\cite{S98}).

Since $\ncr$ is difficult to measure, it is most often assumed that magnetic
fields and cosmic rays are in energy equipartition; this allows one to express
$\ncr$ in terms of $\HH$. The physical basis of this assumption is the fact
that cosmic rays are confined by magnetic fields. An additional assumption
involved is that the energy density of relativistic electrons responsible for
synchrotron emission (and so $\ncr$) is proportional to the total energy
density of cosmic rays; it is usually assumed that the energy density of
relativistic electrons in the relevant energy range (i.e.\ several GeV) is one
percent of the proton energy density in the same energy interval
(Chapter~19 in Longair \cite{L94}). The validity
of this assumption requires that the diffusion/escape losses of the cosmic-ray
electrons dominate over radiative losses, so that the energy spectrum of the
electrons is not steeper than that of the protons (Beck \cite{B97}).

Estimates of $\HH$ in our Galaxy via $\ncr$
determined from $\gamma$-ray emission (Strong et al.\ \cite{S00}) are
generally consistent with the equipartition values (E.~M.~Berkhuijsen, in Beck
\cite{B01}). However, Eq.~(\ref{polar}) is not consistent with the
equipartition or pressure balance between cosmic rays and magnetic fields
insofar as it assumes that $\ncr=\mbox{const}$. Therefore, the regular
magnetic field strength obtained using Eq.~(\ref{polar}) can be inaccurate
(see Section~\ref{EEFSI}).

The thermal electron density $\nel$ in the ISM can be obtained from emission
measure ${\rm EM}\propto\int_L\nel^2\,ds$, although this involves additional
assumptions regarding the filling factor of interstellar clouds. In the Milky
Way, the dispersion measures of pulsars, ${\rm DM}=\int_L\nel\,ds$ provide
information about the mean thermal electron density, but the accuracy is
limited by our uncertain knowledge of distances to pulsars.
Estimates of the strength of the regular magnetic field in the Milky Way are
often obtained from the Faraday rotation measures of pulsars simply as
\begin{equation}        \label{obsB}
\BB_\parallel\simeq\frac{\RM}{K_1\,\rm DM}\;.
\end{equation}
This estimate is meaningful if magnetic field and thermal electron
density are {\em uncorrelated\/}.
If the fluctuations in magnetic field and thermal electron density
are correlated with each other, they will contribute positively to $\RM$
and Eq.~(\ref{obsB}) will yield overestimated $\BB_\parallel$. In the case
of anticorrelated fluctuations, their contribution is negative and
Eq.~(\ref{obsB}) is an underestimate. In order to quantify this effect, one needs
a suitable model for the relation between magnetic fields and thermal
electron density.
As we show in Section~\ref{TEOMIFORM}, physically reasonable
assumptions about the statistical relation between magnetic field strength and
electron density can lead to Eq.~(\ref{obsB}) being in error by a factor of
2--3 even in a statistically homogeneous magneto-ionic medium.
Lerche (\cite{L70}) discussed the effects of
correlated fluctuations in magnetic field and electron density on
Faraday rotation measures. Some results of that
paper are presented in a questionable form,
although the general conclusion agrees with that proposed here.

Equation~(\ref{obsB}) can
also lead to significantly underestimated strength of the
regular magnetic field if it is enhanced in the interarm regions whereas
electron density is maximum within the arms, as in galaxies with magnetic arms
(Beck \cite{B01}).

\section{Magnetic field estimates}        \label{MFE}
The observable quantities (\ref{ints}) provide extensive data on magnetic
field strengths in both the Milky Way and external galaxies (Ruzmaikin et al.\
\cite{RSS88}; Beck et al.\ \cite{BBMSS96}; Beck \cite{B00}, \cite{B01}). The
average total field strengths in nearby spiral galaxies obtained from total
synchrotron intensity $I$ range from $\HH\simeq4\mkG$ in the galaxy M31 to
$\simeq15\mkG$ in M51, with the mean for the sample of 74 galaxies of
$\HH\simeq9\pm3\mkG$ (Beck \cite{B00}). The typical degree of polarization of
synchrotron emission from galaxies at short radio wavelengths is
$p=10\%$--$20\%$, so Eq.~(\ref{polar}) gives $\BB/\HH=0.4$--0.5; these are
always lower limits due to limited resolution of the observations. Most
existing polarization surveys of synchrotron emission from the Milky Way,
having much better spatial resolution, suffer from Faraday depolarization
effects and missing large-scale emission and cannot provide reliable values
for $p$ (see Sect.~\ref{Discussion}). Phillipps et al.\ (\cite{P81}) obtained
$\BB/\HH=0.6$--0.7 from analysis of the total synchrotron emission from the
Milky Way along and perpendicular to the spiral arms. Heiles (\cite{H96})
derived similar values from starlight polarization data. The total
equipartition magnetic field in the Solar neighbourhood is estimated as
$\HH=6\pm2\mkG$ from the synchrotron intensity of the large-scale, diffuse
Galactic radio background (E.~M.~Berkhuijsen, in Beck \cite{B01}). Combined
with $\BB/\HH=0.65$, this yields a strength of the local regular field of
$\BB=4\pm1\mkG$. Hence, the typical strength of the local Galactic random
magnetic fields, $\langle\bb^2\rangle^{1/2}=(\HH^2-{\BB}^2)^{1/2}=5\pm2\mkG$,
exceeds that of the regular field by a factor
$\langle\bb^2\rangle^{1/2}/\BB=1.3\pm0.6$. $\RM$ data yield similar values
for this ratio (Sect.~IV.4 in Ruzmaikin et al.\ \cite{RSS88} ; Ohno \& Shibata
\cite{OS93}).

Meanwhile, the values of $\BB$ in the Milky Way obtained from Faraday rotation
measures seem to be systematically lower than the above values. $\RM$ of
pulsars and extragalactic radio sources yield $\BB=1.4\pm0.3\mkG$ in the local
(Orion) (Rand \& Lyne \cite{RL94}; Frick et al.\ \cite{F01}),
$\BB=1.7\pm0.3\mkG$ in the Sagittarius--Carina spiral arm (Frick et al.\
\cite{F01}); for the Perseus arm, Frick et al.\ (\cite{F01}) obtained
$\BB=1.4\pm1.2\mkG$, and Mitra et al.\  (\cite{M2003}), $\BB=1.7\pm1.0\mkG$.
The median value of $\BB$ is about twice smaller than that inferred from other
methods.

There can be several reasons for the discrepancy between the estimates of the
regular magnetic field strength from Faraday rotation and synchrotron
intensity. Both methods suffer from systematic errors due to our uncertain
knowledge of thermal and relativistic electron densities, so one cannot be
sure if the difference is significant. Nevertheless, the discrepancy seems to
be worrying enough to consider carefully its possible reasons. (We should
emphasize that the main results of this paper are independent of whether or
not the discrepancy is real.)

The discrepancy can be explained, at least in part, if the methods described
above sample different volumes. The observation depth of total synchrotron
emission, starlight polarization and of Faraday rotation measures are all of
the order of a few kpc. Polarized emission, however, may emerge from more
nearby regions (see Sect.~\ref{Discussion}). However, there are more
fundamental reasons for the discrepancy that we discuss in what follows.

\section{The effects of magneto-ionic fluctuations on RM} \label{TEOMIFORM}

In this section we show that a statistical correlation between electron
density fluctuations and turbulent magnetic fields can affect significantly
regular magnetic field estimates obtained from Faraday rotation measures. The
effect of fluctuations in the magneto-ionic medium on $\RM$ can be the main
reason for the discrepancy between magnetic field estimates discussed in
Sect.~\ref{MFE}. Note that the effect discussed here arises from correlations
at scales $\la100\p$ (i.e., less than the basic turbulent scale), rather than
from any relations between the averaged quantities (e.g., arising from the
vertical stratification of the ISM or galactic density waves).

In order to quantify the effect, one needs a specific physical model for the
connection between thermal gas density and  magnetic field. An appealing idea
is based on the assumption that the ISM is in {\em pressure balance\/}
involving not only thermal pressure (Field et al.\ \cite{FGH69}; McKee \&
Ostriker \cite{MO77}) but also turbulent, magnetic and cosmic ray pressures
(e.g., Parker \cite{P79}; Boulares \& Cox \cite{BC90}; Fletcher \& Shukurov
\cite{FS01}). Pressure balance can be maintained at the scales of interest
since turbulence is subsonic,  $v\la\cs$ where $v$ is the turbulent velocity,
and so the sound crossing time $l/\cs$ (across the correlation length of the
turbulent magnetic fields $l$) is comparable to or shorter than the presumed
correlation time of interstellar turbulence $l/v$. Numerical simulations of
the multiphase ISM driven by supernova explosions indicate statistical
pressure equilibrium over a wide range of physical conditions in the ISM, both
between different phases and within a region occupied by a single phase
(Rosen et al.\ \cite{RBK96}; Korpi et al.\ \cite{K99}; Gazol et al.\
\cite{GVSS01}). Anyway, any system that deviates from pressure equilibrium on
average must either expand or collapse; thus, any statistically steady state
of the ISM must involve statistical pressure equilibrium.

Treatment of pressure equilibrium involving magnetic field requires some
caution since magnetic pressure only contributes to force balance across the
field lines. However, splitting magnetic field into an (isotropic) random part
and a large scale field, as we do here, admits the reasonable assumption that
magnetic pressure due to the random part is isotropic as long as we consider
quantities averaged over an intermediate scale which is larger than the
turbulent correlation scale ($l=50$--100\p) but smaller than the scale of the
regular field ($\ga1\kpc$).

The peculiar nature of the Lorentz force still remains important, e.g.\
magnetic buoyancy would produce gas motions even under perfect pressure
balance. However, magnetic buoyancy is important at relatively large scales of
the order of 1\,kpc (Parker \cite{P79}), and so can be neglected here. Indeed,
its characteristic time, equal to the Alfv\'en crossing time over the density
scale height, cannot be shorter than the sound crossing time over the
turbulent correlation length because both the turbulent scale is smaller than
the density scale height and the Alfv\'en speed is smaller than the speed of
sound (again assuming that the turbulence is subsonic). It is therefore not
unreasonable that studies of magnetic buoyancy often start with a
pressure-equilibrium state as an initial condition. Hence, we can neglect any
effects related to anisotropy of the pressure produced by the large-scale
magnetic field.

Pressure equilibrium in the ISM naturally leads to an anticorrelation between
gas density and magnetic fields because regions with enhanced magnetic field
must have lower density (if only the cooling time is longer than the sound
crossing time, so that a transient pressure excess cannot be removed by
cooling). If pressure equilibrium is only maintained in the statistical sense,
the anticorrelation will also be only statistical.

On the other hand, there are systematic deviations from pressure balance in
the ISM. For example, the total pressure is larger than average in
self-gravitating clouds, expanding supernova remnants, stellar wind regions,
and in travelling galactic density waves. Such overpressured regions can
result in a positive correlation between magnetic field and gas density,
leading to enhanced Faraday rotation.

We derive an expression for the Faraday rotation measure under the assumption
of pressure equilibrium in Sect.~\ref{MIMPB} and generalize it in
Sect.~\ref{Over}. Here we employ the assumption of isotropic random magnetic
field. As discussed in Sect.~\ref{AMF}, this assumption is only approximate,
but deviations from statistical isotropy can hardly affect significantly the
effects discussed in this section.

\subsection{Magneto-ionic medium in pressure balance}      \label{MIMPB}
In order to evaluate a correction to $\RM$ resulting from the
anticorrelation of $\bb$ and $\nel$, we first obtain $\nel$ as a
function of $\bb$ from pressure equilibrium equation. We introduce
Cartesian coordinates $(x,y,z)$ with the $x$-axis parallel to
$\vec{\BB}$, so that $\vec{\BB}=(\BB,0,0)$, and the line-of-sight
vector $\vec{s}=(s_x,s_y,s_z)$
directed towards the observer, with $L$  the distance
from the source of polarized emission to the observer. Assuming that
cosmic ray pressure is equal to the magnetic pressure,
we start with pressure equilibrium equation,
\begin{equation}
\frac{\HH^2}{4\pi}+\Pgas={\cal P}\;,           \label{eq}
\end{equation}
where $\Pgas$ is the gas pressure comprising thermal and turbulent
components, the magnetic contribution has been doubled to allow for
cosmic ray pressure, and ${\cal P}$ is the total pressure.
Since $\Pgas$ is proportional to the total gas density, we write
\[
\Pgas=\nel F\;,
\]
where $F=\Pgas/(X\ntot)$ with $\ntot$ the total gas density and $X$ the degree
of ionization. This representation is convenient if the medium is isothermal,
which is a good approximation since $\RM$ is mainly produced in a single phase
of the ISM, the warm ionized medium (Heiles \cite{H76}). A more consistent
theory should include fluctuations in the total pressure
and variable degree of ionization, together with a more
realistic equation of state (e.g., polytropic).  We leave these
generalizations for future work, and only briefly discuss the polytropic
equation of state in Sect.~\ref{Over}.

Consider magnetic field $\HH$ consisting of a regular $\BB$ and
turbulent parts, and similarly for electron density:
\[
\vec{\HH}=\vec{\BB}+\vec{\bb}\;,
\quad
\nel=\N+\n\;,
\quad
\langle\vec{\bb}\rangle=\langle\n\rangle=0\;;
\]
we shall assume that
the random field $\vec\bb$ is isotropic.
Averaging Eq.~(\ref{eq}) yields
\begin{equation}
\BB^2+\langle\bb^2\rangle=4\pi({\cal P}-\N F)\;,   \label{Bb}
\end{equation}
where we have assumed that $\cal P$ and $F$ do not fluctuate. Subtract
(\ref{Bb}) from (\ref{eq}) to obtain the following balance equation for the
fluctuations:
\begin{equation}
\bb^2-\langle\bb^2\rangle+2\vec{\BB}\cdot\vec{b}=-4\pi\n F\;,
                                                        \label{fluc}
\end{equation}
or
\begin{equation}			\label{nnn}
\n=-\frac{1}{4\pi F}(\bb^2-\langle\bb^2\rangle)
        -\frac{1}{2\pi F}\vec{\BB}\cdot\vec{\bb}\;.\
\end{equation}
This equation shows that electron density is smaller ($\n<0$) where
magnetic field is stronger, because either $\bb^2$ is larger than
average or $\vec{\BB}$ and $\vec{\bb}$ are similarly directed,
$\vec{\BB}\cdot\vec{\bb}>0$.

Now we calculate $\RM$ identifying the integral in Eq.~(\ref{ints}) with
ensemble average:
\begin{equation}
\RM/K_1L=\langle\vec{\BB}\cdot\vec{s}\nel\rangle
=(\vec{\BB}\cdot\vec{s})\N+\langle(\vec{\bb}\cdot\vec{s})\n\rangle\;;
\label{RM}
\end{equation}
where $\vec{s}$ is the line-of-sight vector.
In order to calculate the last term on the right-hand side, we use
Eq.~(\ref{nnn}) to obtain
\begin{equation}          \label{nbs}
\langle\n\vec{\bb}\cdot\vec{s}\rangle
=-\frac{1}{2\pi F}
        \langle(\vec{\bb}\cdot\vec{s})(\vec{\BB}\cdot\vec{\bb})\rangle\;,
                        \label{eee}
\end{equation}
since
$\left\langle(\vec{\bb}\cdot\vec{s})\left(\bb^2-\langle\bb^2\rangle\right)\right\rangle=0$
because $\langle\vec{\bb}\cdot\vec{s}\rangle=0$ and
$\langle(\vec{\bb}\cdot\vec{s})b^2\rangle
=\langle b^3\cos\alpha\rangle=0$ (with $\alpha$ the
random angle between $\vec\bb$ and $\vec{s}$) as $\vec{\bb}$ is an
isotropic random vector.

The average in Eq.~(\ref{eee}) can be calculated expanding the dot products
into Cartesian components:
\begin{eqnarray*}
\langle(\vec{\bb}\cdot\vec{s})(\vec{\BB}\cdot\vec{\bb})\rangle &=&
\langle(\bb_xs_x+\bb_ys_y+\bb_zs_z)(\BB\bb_x)\rangle\\
&=&\BB s_x\langle\bb_x^2\rangle+\BB s_y\langle\bb_x\bb_y\rangle
                +\BB s_z\langle\bb_x\bb_z\rangle\;.
\end{eqnarray*}
We assume that $\bb_x$, $\bb_y$ and $\bb_z$ are uncorrelated when evaluated at
the same position, so $\langle\bb_xb_y\rangle=\langle\bb_x\bb_z\rangle=0$.
This condition is not restirctive as it is satisfied by any random magnetic
field  $\vec{\bb}$ with symmetric probability distributions of $\bb_x,\ \bb_y$
and $\bb_z$ (and for an isotropic $\vec{\bb}$ in particular). Then
\begin{equation}                        \label{bsbb}
\langle(\vec{\bb}\cdot\vec{s})(\vec{\BB}\cdot\vec{\bb})\rangle
                        =\sfrac13\BB_\parallel \langle\bb^2\rangle\;,
\end{equation}
where $\BB_\parallel= \vec{\BB}\cdot\vec{s}\equiv \BB s_x$ is the
line-of-sight regular magnetic field and
$\sfrac13\langle\bb^2\rangle=\langle\bb_x^2\rangle$ due to the isotropy
of $\vec{\bb}$.

We note that the assumption that the components of
$\vec{\bb}$ are statistically point-wise independent
is compatible with the solenoidality of magnetic field because the latter
only requires that magnetic field components are correlated when taken at
distinct positions. In particular, the components of a random, isotropic
and homogeneous magnetic field must be statistically independent when
taken at the same position.

Using Eqs (\ref{eee}) and (\ref{RM}), we obtain
\begin{eqnarray}   \label{res1}
\RM&=&K_1\left[\BB_\parallel\N L-\frac{1}{6\pi F}\BB_\parallel L
\langle\bb^2\rangle\right]	\nonumber\\
&=&\RM^{(0)}                                  \label{res0}
\left(1-\sfrac23\,\frac{\langle\bb^2\rangle}{4\pi\langle\Pgas\rangle}\right)\;,
\end{eqnarray}
where $\RM^{(0)}=K_1\BB_\parallel \N L$ is the Faraday rotation
measure that would arise if magnetic field and electron density
fluctuations were uncorrelated. So, $\RM$ has a negative contribution
from the turbulent magnetic field. We can rewrite this formula
in terms of magnetic field alone assuming
that the average magnetic pressure is equal to the gas pressure
(equipartition between magnetic and gas energies),
\begin{equation}                \label{equip}
\BB^2+\langle\bb^2\rangle=4\pi\langle\Pgas\rangle\;,
\end{equation}
which yields
\begin{equation}			\label{res}
\RM=\RM^{(0)}
\left(1-\sfrac23\,\frac{\langle\bb^2\rangle}{\BB^2+\langle\bb^2\rangle}\right)\;.
\end{equation}
The equipartition expressed by Eq.~(\ref{equip}) is
assumed to hold at scales  $\ga1\kpc$, so that it does not
contradict the assumption of pressure equilibrium (Eq.~(\ref{eq})) that holds
at small scales, $\la1\kpc$.

   \begin{figure}
   \centering
   \includegraphics[width=0.45\textwidth]{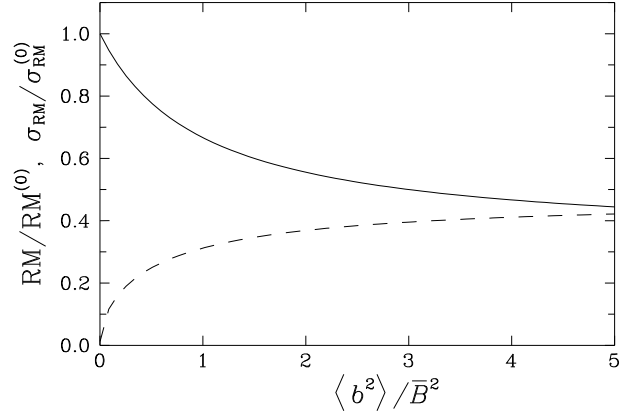}
      \caption{The ratio of $\RM$ (solid) and $\sigma_{\rm RM}$
(dashed) to their standard estimates $\RM^{(0)}$ and
$\sigma_{{\rm RM}^{(0)}}$ as given by Eqs~(\protect\ref{res}) and
(\protect\ref{sRM}), respectively. In Eq.~(\protect\ref{sRM}),
we have taken $f=0.1$ and
$\BB_\perp^2=\BB_\parallel^2=\sfrac12\BB^2$.}
\label{fig1}
\end{figure}

The dependence of $\RM/\RM^{(0)}$ on $\langle\bb^2\rangle/\BB^2$ is shown in
Fig.~\ref{fig1}. In the limiting case $\langle\bb^2\rangle/\BB^2\gg1$, we have
\[
\RM=\sfrac13\RM^{(0)}\;,
\]
i.e., $\BB$ would be  underestimated by a factor of 3 if calculated as
$\BB=\RM/K_1\N L$ as usually done. As discussed in Sect.~\ref{MFE}, the
relative strength of the random magnetic field in the Milky Way is
$\langle\bb^2\rangle^{1/2}/\BB=1.3\pm0.6$; we adopt
$\langle\bb^2\rangle/\BB^2=2$ as a representative value. Then Eq.~(\ref{res})
shows that the standard estimate of $\BB$ is about two times too small. In
other words, the regular magnetic field strength near the Sun consistent with
the Faraday rotation measures of pulsars and extragalactic radio sources is
$\BB\simeq3$--$4\mkG$ (twice the value usually inferred from $\RM$ data), in
agreement with the estimates from equipartition between cosmic rays and
magnetic fields.

Apart from the reduction in the observed $\RM$, quantified by Eq.~(\ref{res}),
if there were an anticorrelation between magnetic field and electron
density fluctuations, then the observed fluctuations in Faraday rotation
measure would be reduced as well.  When fluctuations in
magnetic field and electron density are uncorrelated, the standard deviation
of $\RM$ is given by
\begin{equation}                        \label{sRM0}
\sigma_{\RM}^{(0)}
=K_1\left(2\langle n^2\rangle \langle b^2\rangle Ld\right)^{1/2},
\end{equation}
where $d$ is the common correlation length of magnetic and electron
fluctuations (see, e.g., Appendix~A in Sokoloff et al.\ \cite{S98}). As shown
in Appendix~\ref{sigma}, under conditions where Eq.~(\ref{res}) applies, the
standard deviation of $\RM$
is given by
\begin{eqnarray}                        \label{sRM}
\frac{\sigma_{\RM}}{ \sigma_{\RM}^{(0)}}&=&
\sfrac{40}{9\pi} \left(\frac{f}{1-f}\right)^{1/2}
\frac{\langle b^2\rangle}{\BB^2+\langle b^2\rangle} \nonumber\\
&&\mbox{}\times\left(1+\sfrac{9}{20}\,\frac{\BB_\perp^2}{\langle b^2\rangle}
-\sfrac{3}{40}\,\frac{\BB_\parallel^2}{\langle b^2\rangle}\right)^{1/2},
\end{eqnarray}
where $f$ is the filling factor of thermal electrons, defined as
$f=N^2/\langle\nel^2\rangle$. (Of course numerical coefficients in this
formula are model dependent.) Since $f\ll1$
(Berkhuijsen \cite{B99} and references therein), the anticorrelation suppresses
fluctuations in Faraday rotation, $\sigma_{\RM}/\sigma_{\RM}^{(0)}\la1$. This
can be clearly seen in Fig.~\ref{fig1}. For example,
$\sigma_{\RM}/\sigma_{\RM}^{(0)}\simeq0.4$ for $\langle\bb^2\rangle/\BB^2=2$
and $f=0.1$ (Berkhuijsen \cite{B99}). In the limiting case of strong magnetic
fluctuations, $\langle\bb^2\rangle/\BB^2\gg1$, the right-hand side of
Eq.~(\ref{sRM}) is approximately equal to $f^{1/2}$. An important result of
the reduction of the standard deviation of the Faraday rotation measure is
that this reduces the depolarization effect of internal Faraday dispersion in
comparison with standard estimates. 

Another implication is that the value of $\sigma_{\RM}$ depends on the
direction, it is minimum along the regular magnetic field and maximum in the
directions across it. We stress that this anisotropy in $\sigma_{\RM}$ arises
even in an isotropic random magnetic field. Brown \& Taylor (\cite{BT01}) find
that the scatter in $\RM$ is larger along the Orion arm than across it,
contrary to our result. As we discuss in Sect.~\ref{AMF}, it is natural to
expect that the turbulent magnetic field in the ISM is anisotropic in a manner
consistent with the observations of Brown \& Taylor, and that the anisotropy
in $\vec{\bb}$ masks the more subtle effect discussed in this section.

\subsection{Overpressured regions and the polytropic equation of
state} \label{Over}

Systematic and random deviations from pressure equilibrium are widespread in
the ISM, and this affects the relation between local magnetic field and
electron density. For example, compression leads to enhancement in both
magnetic field and gas density, and these variables
{\bf can} be
correlated. (We note that the anticorrelation discussed in Sect.~\ref{MIMPB}
is difficult to detect because observational estimates are biased towards
dense regions.) In this section we discuss the effect of the positive
correlation between $\bb$ and $\nel$ at scales $\la100\p$, i.e., where the
correlation (if any) can have a random character;
such regions are not likely to be widespread in the diffuse ISM.
As might be expected, this leads to
enhanced Faraday rotation. Next we consider the polytropic equation of state,
where the assumption of isothermal medium of Sect.~\ref{MIMPB} is relaxed.

Suppose that the total magnetic field scales with the total electron
density as
\begin{equation}
\frac{\HH^2}{4\pi} =a \nel^\kappa
= a(\N+n)^\kappa \simeq a(\N^\kappa + \kappa n \N^{\kappa-1})\;,
\label{eeq}
\end{equation}
where
$a$ and $\kappa$ are constants.
The last equality is
only valid for weak fluctuations, $n/N \ll 1$. Averaging Eq.~(\ref{eeq})
yields
\begin{eqnarray}
\BB^2+\langle\bb^2\rangle&=&4\pi a\N^\kappa\;,         \label{BBb}\\
\bb^2-\langle\bb^2\rangle+2\vec{\BB}\cdot\vec{\bb}&
=&4\pi a\kappa \n \N^{\kappa-1}\;,                     \label{ffluc}
\end{eqnarray}
which are useful to compare with Eqs~(\ref{Bb}) and (\ref{fluc}),
respectively.  Calculations similar to those in Sect.~\ref{MIMPB}
(with $F$ replaced by $a\kappa N^{\kappa-1}$) then yield
\begin{equation}                        \label{rres}
\RM=\RM^{(0)}
\left(1+\sfrac23\kappa\,\frac{\langle\bb^2\rangle}{\BB^2+\langle\bb^2\rangle}\right)\;,
\end{equation}
which can be compared with Eq.~(\ref{res}). Thus, a positive
(negative) correlation between $\HH$ and $\nel$, i.e., $\kappa>0$
($\kappa<0$) results in a positive (negative) bias in the observed
$\RM$.
For example, we get $\RM=\sfrac53\RM^{(0)}$
for $\kappa=1$ and $\langle\bb^2\rangle/\BB^2\gg1$,
i.e., the standard estimate of $\BB$ is by a factor 5/3 too large.

The corresponding effect on the standard deviation of $\RM$ is an additional
factor $\kappa$ on the right-hand side in Eq.~(\ref{sRM}).

Now consider a polytropic equation of state, $\Pgas = \nel^\gamma F_1$, where
the factor $F_1$ is similar to $F$ in Sect.~\ref{MIMPB}. In order to simplify
the calculations, we assume that density fluctuations are weak, $\n\ll\N$, so
that $\nel^\gamma\simeq\N^\gamma+\gamma\n\N^{\gamma -1}$. Averaging
Eq.~(\ref{eq}) now yields
\[
\BB^2 + \langle\bb^2\rangle \simeq 4\pi(P-\gamma n N^{\gamma-1} F_1)\;,
\]
and instead of Eq.~(\ref{res}) we obtain Eq.~(\ref{rres}) with $\kappa$
replaced by $-\gamma$.

\section{Magnetic field estimates from synchrotron intensity}\label{EEFSI}

In this section we discuss how anisotropy of the interstellar random magnetic
field and the dependence of the cosmic ray number density on magnetic field
affect the estimates of the regular magnetic field from synchrotron intensity.

\subsection{Anisotropic magnetic fields}        \label{AMF}

Equation~(\ref{polar}) applies only to isotropic random magnetic fields, but
their anisotropy  results in stronger polarization for a given $\bb/\BB$
(Laing \cite{L81,L02}; Sokoloff et al.\ \cite{S98}). Galactic differential
rotation can make the {interstellar turbulent} magnetic field anisotropic by
extending turbulent cells along azimuth so that their azimuthal and radial
sizes are related via $l_\phi/l_r\simeq1+\Delta\Omega R\Delta t\simeq1.3$ at
$R=R_\odot$, where $\Delta\Omega$ is the angular velocity increment across the
turbulent cell, $R$ is the galactocentric radius with $R_\odot$ its Solar
value, and $\Delta t\simeq10^7\yr$ is the lifetime of a turbulent eddy. This
produces anisotropy in the turbulent magnetic field, with
$\bb_\phi/\bb_r\simeq l_\phi/l_r$.

A more fundamental reason of anisotropy in $\vec{\bb}$ is the anisotropic
nature of magnetohydrodynamic turbulence (see Goldreich \& Sridhar
\cite{GS97} and references therein), where turbulent `cells' are elongated
along the mean magnetic field (Goldreich \& Sridhar \cite{GS95}), i.e. roughly
in the azimuthal direction in the case of Galactic magnetic fields.

Anisotropy of both types has the largest component of the random magnetic
field roughly elongated with the large-scale Galactic magnetic field
$\bb_\phi\ga\bb_r\simeq\bb_z$. This is similar to the anisotropy found by
Brown \& Taylor (\cite{BT01}) in the fluctuations in the Galactic $\RM$.

The degree of polarization produced by the above anisotropic random
magnetic field when observed
in directions close to the Galactic centre (where
$\vec{\bb}_\perp$ consists mainly of $\bb_\phi$ and $\bb_z$ with
$\bb_z\simeq\bb_r$) is given by [see Eq.~(19) in Sokoloff et al.\
\cite{S98}]

\begin{equation}                \label{polaranis}
p\simeq p_0
\frac{\BB_\perp^2+\sfrac13\langle\bb^2\rangle(a^2-1)}%
{\BB_\perp^2+\sfrac13\langle\bb^2\rangle(a^2+1)}\;,
\end{equation}
where $a=\bb_\phi/\bb_r$ is a measure of anisotropy. For $a=1$, this
equation reduces to Eq.~(\ref{polar}).
This estimate also applies to directions perpendicular to the spiral arms as
long as their pitch angle is small.

\begin{figure}
\centering
\includegraphics[width=0.45\textwidth]{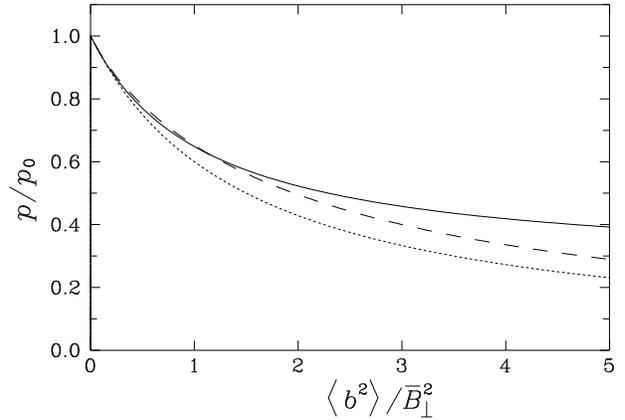}
\caption{
The fractional polarization with anisotropic random magnetic
field as given by Eq.~(\ref{polaranis}) with $a=1.3$ (solid),
under the local equipartition between cosmic rays and magnetic field
(dashed) according to Eq.~(\ref{polareq}), and from the standard relation
(\ref{polar}) (dotted) as functions of the ratio
$\langle \bb^2 \rangle / \BB_\perp^2$.
The standard relation (\ref{polar}) gives significantly overestimated
$\BB_\perp$
for a given fractional polarization $p$ and random field
strength
$\langle\bb^2\rangle^{1/2}$ if
$\langle \bb^2 \rangle / \BB_\perp^2>2$.}
\label{fig2}
\end{figure}

The solid line in Fig.~\ref{fig2} shows the fractional
polarization as given by Eq.~(\ref{polaranis}); it is significantly
different from that given by Eq.~(\ref{polar}) for realistic values
of parameters.  The same value of $p$ in fact corresponds to a larger
value of $\langle\bb^2\rangle/\BB_\perp^2$ when the anisotropy
has been taken into account. Since the energy density
in the random magnetic field is limited from above by the kinetic
energy of interstellar turbulence, this implies that values of the
regular magnetic field strength obtained from Eq.~(\ref{polar}) are
{\em overestimated\/}.  For example, the degree of polarization of
$p/p_0=0.4$ is obtained for
$\langle\bb^2\rangle/\BB_\perp^2\simeq3.0$ according to
Eq.~(\ref{polaranis}) and for
$\langle\bb^2\rangle/\BB_\perp^2\simeq2.2$ according to
Eq.~(\ref{polar}). For a fixed $\langle\bb^2\rangle$, the difference
in $\BB_\perp$ is about 1.2.

Thus, because of the anisotropy of the turbulent magnetic field with
$\bb_\phi\ga\bb_r\simeq\bb_z$, the regular magnetic field obtained from
polarized intensity can be significantly overestimated
(whereas estimates of the {\it total} field remain unaffected). In the Milky Way, this
effect is strongest in, roughly, the azimuthal direction where
$\vec{\bb}_\perp$ is dominated by $\bb_\phi$ and $\bb_z$ and weakest toward
the Galactic centre (nearly across the spiral arms) where the main
contribution to $\vec{\bb}_\perp$ comes from $\bb_r$ and $\bb_z$. The effect
of the anisotropy is undoubtedly significant in external galaxies.

\subsection{Local equipartition between cosmic rays and magnetic fields} \label{LE}

Arguments similar to those of Sect.~\ref{TEOMIFORM} can be applied to cosmic
ray density that may exhibit similar local dependence on the magnetic field
strength, with similar consequences for the magnetic field estimates from the
total and polarized synchrotron intensities. If the energy equipartition or
pressure balance between cosmic rays and magnetic fields were maintained
locally (i.e., at any given position), one would expect a strong positive
correlation between $\ncr$ and $\HH$ (given that $\ncr$ is proportional to the
total cosmic ray energy density). Sarkar (\cite{S82}) argues that such a
correlation, implied by cosmic ray confinement ideas, can be enhanced by
acceleration of relativistic electrons in compressed supernova shells.

In fact, Eq.~(\ref{polar}) widely used in interpretations of polarized radio
emission from the Milky Way and external galaxies is inconsistent with energy
equipartition or pressure balance between cosmic rays and magnetic fields
because it is based on the assumption that the number density of cosmic rays
is independent of magnetic fields. This also applies to Eq.~(\ref{polaranis}).
If the local equipartition is maintained between cosmic rays and magnetic
fields, so that $\ncr\propto\HH^2$ at any position, the total intensity of
synchrotron emission will strongly depend on the magnetic field since
synchrotron emissivity is proportional to
$\ncr\HH_\perp^{1+\alpha} \propto\HH^2\HH_\perp^{1+\alpha}$, where
the synchrotron spectral index $\alpha$ is close to unity.

As a result, the
equipartition estimate will be biased towards regions with stronger field and
the degree of polarization will be larger than predicted by Eq.~(\ref{polar}).
Sokoloff et al.\ [\cite{S98}, their Eq.~(28)] have shown that for
$\ncr\propto\HH^2$ Eq.~(\ref{polar}) is replaced by
\begin{equation}        \label{polareq}
p = p_0
\frac{\BB_\perp^2+\sfrac73\langle\bb^2\rangle}%
{\BB_\perp^2+3\langle\bb^2\rangle+\sfrac{10}{9}\langle\bb^2\rangle^2} \;,
\end{equation}
where it is assumed for simplicity that the regular magnetic field lies in the
sky plane, $\BB=\BB_\perp$, and the random magnetic field is isotropic. The
resulting degree of polarization is shown in Fig.~\ref{fig2}. As expected,
Eq.~(\ref{polareq}) yields significantly stronger fractional polarization for
a given ratio $\langle\bb^2\rangle/\BB_\perp^2$. For example, for $p/p_0=0.4$,
Eq.~(\ref{polar}) predicts that $\langle\bb^2\rangle/\BB_\perp^2\simeq2.2$,
whereas Eq.~(\ref{polareq}) yields $\langle\bb^2\rangle/\BB_\perp^2\simeq4.7$.
For a fixed strength of the random magnetic field, this leads to an
overestimate by a factor of 1.4 in the regular field strength.
This difference is very
close to that between the estimates obtained from synchrotron intensities and
$\RM$.
The equipartition value of the {\it total} field is also
an overestimate.

A tight, point-wise correlation between $\ncr$ and $\HH$ can be an
oversimplification. Cosmic ray diffusion can smooth variations in $\ncr$ at
fairly large scales. Then Eq.~(\ref{polareq}) provides an upper estimate of
$p$. Otherwise, one can still use Eq.~(\ref{polareq}), but with
$\langle\bb^2\rangle$ understood as an average over the diffusion length of
cosmic-ray electrons.

There are other uncertainties in the equipartition estimates of field
strengths. For example, the standard estimate relies on the assumption that
relativistic protons and electrons have the same spectral index in the
relevant part of the energy spectrum (i.e. around a few GeV for electrons
responsible for radio synchrotron emission). This is true in the Solar
neighbourhood (Chapter~9 in Longair \cite{L94}). However, this is no longer
valid if the electrons suffer significant synchrotron or inverse Compton
losses, so that their energy spectrum is steeper than that of the protons.
In this case the equipartition or minimum-energy method should be applied
with care and requires a correction of the electron spectrum derived
from the synchrotron spectrum.

\begin{table*}
\caption[]{\protect\label{results}
Summary of the various effects of random magnetic fields on Faraday rotation
measure $\RM$ and the degree of synchrotron polarization $p$ discussed in this
paper. Consequences for the `standard' estimates of $\BB$ from the observed
$\RM$ or $p$ are indicated in column 2. Reference to relevant equations is
given in the third column; the text should be consulted for conditions under
which the equations have been obtained; they
are briefly summarised in Column 4, but the
effects themselves are more general. Here $\nel$ and $\ncr$ are the number
densities of thermal and cosmic-ray electrons, respectively,
with $\N=\langle\nel\rangle,$ and $\n=\nel-\N$; $\vec{\HH}$,
$\vec{\BB}$ and $\vec{\bb}$ are the total, regular and random magnetic fields,
respectively.}
\begin{center}
\begin{tabular}{@{}llll@{}}
\hline
Physical effect                     &Consequences
              &Equation                         &Assumptions\\
&             &reference\\
\hline
$\nel$--$\HH$ anticorrelation &$\RM$ reduced, $\BB$ underestimated, $\vphantom{{B^2}^2}$
              &(\ref{res0})                     &pressure balance, isotropic $\vec{\bb}$\\[3pt]
     &        &(\ref{res})                      &$+$ magnetic and turbulent energies in equipartition\\[3pt]
                              &$\sigma_{\RM}$ reduced &(\ref{sRM}) &as for (\ref{res})\\[3pt]
$\nel$--$\HH$ correlation     &$\RM$ enhanced, $\BB$ overestimated
              &(\ref{rres})                     &isotropic $\vec{\bb}$, $\n/\N\ll1$\\[3pt]
Anisotropy of $\vec{\bb}$           &$p$ enhanced, $\BB$ overestimated
              &(\ref{polaranis})                &$\ncr$ independent of $\BB$\\[3pt]
$\ncr$--$\HH$ correlation           &$p$ enhanced, $\BB$ and $\HH$ overestimated
              &(\ref{polareq})                  &isotropic $\vec{\bb}$\\[3pt]
\hline
\end{tabular}
\end{center}
\end{table*}

\section{Discussion} \label{Discussion}

In order to verify our main results, Eqs.~(\ref{res}) and (\ref{rres}), one
would need to study the relation between $\RM$, $P$ and $p$ because
Eqs.~(\ref{res}) and  (\ref{rres}) can be rewritten as
\begin{equation}
\RM\simeq \RM^{(0)}               \label{rres2}
\left[1+\sfrac23\kappa\left(1-\sfrac23\,\frac{p}{p_0}\right)\right]\;,
\end{equation}
where Eq.~(\ref{res}) corresponds to $\kappa=-1$, and Eq.~(\ref{polar}) has
been used with $\BB_\perp^2=\sfrac12\BB^2$ and $p\ll p_0$. In other words,
statistical relation between magnetic field and electron density fluctuations
leads to a correlation of the observable Faraday rotation measure with the
degree of polarization. In order to rewrite Eq.~(\ref{rres2}) in terms of
observable quantities, we note that $\BB_\parallel$ can be related to the
observed polarized intensity given in Eq.~(\ref{ints}) since $\BB_\parallel$
and $\BB_\perp$ are related via a geometric factor. Then $\RM^{(0)}$ can be
expressed in terms of polarized intensity $P$, $\RM^{(0)}\propto P^{1/2}$, and
Eq.~(\ref{rres2}) written as
\begin{equation}        \label{obs}
\frac{|\RM|}{\sqrt{P}} \propto 1-\frac{2}{3}\,\frac{1}{1+3/(2\kappa)}\,\frac{p}{p_0}\;,
\end{equation}
where  the coefficient of proportionality depends on cosmic ray number
density, mean thermal electron density, geometric factors, path length and
other variables. Regions with pressure balance yield $1+3/(2\kappa)<0$, and so
a positive correlation between $|RM|/\sqrt P$ and $p$,
whereas overpressured
regions should be detectable via an anticorrelation between these quantities.
Without any interdependence between magnetic field and electron density,
$|\RM|/\sqrt{P}$ would be uncorrelated with the degree of polarization $p$.

In order to detect the correlation implied by Eq.~(\ref{obs}), the
observational data have to be selected carefully.
Regions in which synchrotron emission and Faraday rotation occur together,
so that they probe the same magnetic field, have to be found.
The regular magnetic field must be almost uniformly directed throughout
the region to reduce the variation in the ratio $\BB_\parallel/\BB_\perp$.
The positive and negative
correlations from regions in pressure equilibrium and overpressured
regions can compensate each other. Thus, to detect the correlation
produced by pressure balance, the region must not be affected by any
violent activity that could produce overpressure, e.g., expanding
supernova shells
and  \ion{H}{ii} regions (Jenkins \& Tripp \cite{JT01}).
The region must be very well explored in several
wavelength ranges in order to have reliable Faraday rotation
measures, polarized and total synchrotron intensities, and ionized
and neutral gas densities. Finally, Faraday depolarization must not
be strong to ensure that $P$ is not affected by variations in Faraday
depth.

There are several surveys of the polarized radio background in the Milky Way
(at $\lambda=21$--$74\cm$ -- Spoelstra \cite{S84};
$\lambda=21\cm$ -- Uyan{\i}ker et al.\ \cite{U99};
$\lambda=80$--$88\cm$ -- Haverkorn et al.\ \cite{H00}, \cite{H03}; and
$\lambda=21\cm$ -- Gaensler et al.\ \cite{G01}), but they cover regions
and/or wavelengths where Faraday depolarization is strong and so Faraday
rotation and synchrotron emission occur at different depths; this makes it
difficult to obtain the value of $p$ required. Polarization intensity data at
smaller wavelengths are available (e.g., Duncan et al.\ \cite{D97}), but not
the Faraday rotation. Further away from the Galactic plane, a few regions have
been observed in radio polarization with high resolution, but Faraday rotation
data are available only in small regions (Reich \& Uyan{\i}ker, in prep.).
Further radio polarization studies of carefully selected regions in our Galaxy
are required; Galactic polarization surveys at $\lambda6\cm$ are in planning.

Polarization maps at $\lambda\le6\cm$ are available for several external
galaxies (Beck \cite{B00}). However, the main obstacle is the integration
along the long line of sight (several kpc) and the large beam (several 100~pc)
that mixes a range of heights above the disc and/or positions within and
outside the arms.  The available resolution of observations is insufficient to
isolate relatively small regions in the plane of the sky where pressure
balance can be expected to occur.

An alternative can be the verification of the effect on the standard deviation
of $\RM$, described by Eq.~(\ref{sRM}). A confirmation of the diminishing
effect of pressure equilibrium on fluctuations in the Faraday rotation measure
of pulsars may have been provided by the results of Mitra et al.\
(\cite{M2003}) who have shown that the deviations of pulsar Faraday rotation
measures from a large-scale model distribution are systematically smaller than
predicted by Eq.~(\ref{sRM0})
(we note that the `estimated $\sigma_{\rm rm}$' in their Fig.~8
can be misleading as this quantity is obtained by averaging
residuals squared, and so dominated by a few strong deviations).
The difference appears to be close to a factor of two, which is consistent
with $\langle\bb^2\rangle/\BB^2\simeq3$ (see Fig.~\ref{fig1}).

The apparent disagreement between the values of the regular magnetic field in
the Milky Way, obtained from Faraday rotation measures and synchrotron
intensity, can be resolved by allowing for the effects of turbulent magnetic
fields, and their correlations with thermal and relativistic electron
densities. A detailed comparison of the various methods to obtain field
estimates (radio polarization, starlight polarization, extragalactic and
pulsar rotation measures) in a selected field of the Milky Way is required to
derive reliable estimates of the regular magnetic field in the Milky Way.
Regarding external galaxies, estimates of the regular magnetic field should be
reconsidered with allowance for the anisotropy of the random magnetic field
and with more consistent implementation of the cosmic ray equipartition
models.

An improved estimate of the strength of the regular magnetic field in the
Milky Way can be important, among many other topics, for the studies of the
origin of high-energy cosmic rays (Kalmykov \& Khristiansen \cite{KK95}).

\section{Conclusions}
Our main results are summarized in Table~\ref{results}. We have shown that an
anticorrelation between the number density of thermal electrons $\nel$ and
magnetic field strength (to which both random and regular magnetic fields
contribute) reduces the Faraday rotation measure $\RM$. We have quantified
this effect by assuming that the ISM is under pressure equilibrium and shown
that values of the regular magnetic field strength $\BB$ obtained from
relations similar to Eq.~(\ref{obsB}) can be underestimated by a factor of
about 2. This effect alone can explain the systematic difference between the
values of $\BB$ in the Milky Way obtained from Faraday rotation measures and
the degree of synchrotron polarization. Another consequence of the
anticorrelation is a suppression of fluctuations in $\RM$.

On the contrary, a positive correlation between $\nel$ and magnetic field
strength results in enhanced $\RM$ and overestimated $\BB$. This effect can be
important along selected lines of sight in the Milky Way where the
contribution of overpressured regions is significant.

We have also discussed effects that might affect the value of the regular
magnetic field inferred from the degree of polarization of synchrotron
emission $p$. The anisotropy of the random magnetic field in the ISM can
result from the stretching of turbulent cells by Galactic differential
rotation, in addition to a similar anisotropy inherent to magnetohydrodynamic
turbulence. The anisotropy of $\vec{\bb}$  enhances $p$ and can thereby result
in $\BB$ overestimated by a few tens of percent. A correlation between the
energy density of cosmic ray electrons and magnetic field strength also
enhances the degree of polarization. We have shown that equipartition between
cosmic rays and magnetic fields can lead to an overestimate of $\BB$ by
roughly the same amount. This implies that the ratio of turbulent to regular
magnetic field strengths obtained from the observed degrees of synchrotron
polarization should be revised towards larger values;
$\langle\bb^2\rangle/\BB^2\simeq3$--4 seems to be a plausible estimate.

There are further effects on the equipartition estimates of magnetic
field strengths; we briefly discussed the role of energy losses of
cosmic-ray electrons in Sect.~\ref{LE}.

Altogether, the effects discussed in this paper, especially the bias in the
estimates of $\BB$ from $\RM$, can reconcile the estimates of the regular
magnetic field from Faraday rotation measures and polarized synchrotron
emission. However, this requires a more careful interpretation of
observational data, especially recent observations of diffuse synchrotron
emission in the Milky Way and new determinations of Faraday rotation measures
of extragalactic radio sources and pulsars.

\begin{acknowledgements}
We are grateful to D.~Mitra, W.~Reich and M.~Wol\-leben for useful discussions
and to C.~Heiles for critical reading of the manuscript and helpful
comments. This work was supported by the NATO collaborative research grant
PST.CLG 974737, the PPARC Grant PPA/G/S/2000/00528 and the RFBR grant
01-02-16158.
\end{acknowledgements}



\appendix

\section{The standard deviation of RM}        \label{sigma}
Here we derive Eq.~(\ref{sRM}) for the standard deviation of the Faraday rotation
measure, using the relation $\sigma_X^2=\langle X^2\rangle-\langle X\rangle^2$
for the standard deviation of a random variable $X$. Thus, apart from
Eq.~(\ref{nbs}), we have to calculate $\langle(\n\vec{\bb\cdot s})^2\rangle$.
  From Eq.~(\ref{nnn}), we have
\begin{eqnarray*}
4\pi^2 F^2 (\n\vec{\bb\cdot s})^2&=&
        \left[\sfrac14\left(\bb^2-\langle\bb^2\rangle\right)^2\right.\\
        &&\!\!\!\!\!\!\!\mbox{}+\left.\left(\bb^2-\langle\bb^2\rangle\right){\vec{\BB\cdot\bb}}
        +(\vec{\BB\cdot\bb})^2\vphantom{\sfrac14}\right](\vec{\bb\cdot s})^2\;.
\end{eqnarray*}
The mean values of those terms that contain odd powers of $\bb$ vanish because
(i) the components of $\vec\bb$ are statistically point-wise independent
(so that, say,
$\langle\bb_x^3\bb_y^2\rangle=\langle\bb_x^3\rangle\langle\bb_y^2\rangle$) and
(ii) each of the components has a symmetric probability distribution (so, e.g.,
$\langle\bb_x^3\rangle=0$). Therefore,
\[
\left\langle\left(\bb^2-\langle\bb^2\rangle\right)({\vec{\BB\cdot\bb}})
        (\vec{\bb\cdot s})^2\right\rangle=0\;.
\]
The remaining averages are calculated similarly to the following:
\[
\langle\bb^4 (\vec{\bb\cdot s})^2\rangle =\sum_i \langle\bb^4 \bb_i^2\rangle s_i^2
=\langle\bb^4 \bb_x^2\rangle\;,
\]
where we have taken advantage of the isotropy of $\vec{\bb}$ and the fact
that $\vec{s}$ is a unit vector. Now we represent $\bb^4$ in terms of its
Cartesian components to obtain
\[
\langle\bb^4 (\vec{\bb\cdot s})^2\rangle =
\langle\bb_x^6\rangle +\sfrac{16}{27}\langle\bb^2\rangle^3
+\sfrac{50}{3}\langle\bb^2\rangle \langle\bb_x^4\rangle\;,
\]
where we used the isotropy of $\vec{\bb}$, i.e.,
$\langle\bb_x^2\rangle =\sfrac13\langle\bb^2\rangle$.
The higher moments of $\bb_i$ are calculated
assuming that $\bb_i$ is a Gaussian random variable, e.g.,
$\langle\bb_i^6\rangle=\sfrac59\langle\bb^2\rangle^3$
and $\langle\bb_i^4\rangle=\sfrac13\langle\bb^2\rangle^2$.

Then Eq.~(\ref{sRM}) follows after we note that $\BB_\parallel=\BB s_x$,
$\BB^2=\BB_\perp^2+\BB_\parallel^2$, and
\[
\sigma_{\RM}^2=\frac{2K_1Ld}{4\pi^2 F^2}\left[
                \left\langle(\n\vec{\bb\cdot s})^2\right\rangle
-\sfrac19\BB_\parallel^2 \langle\bb^2\rangle
\right]\;,
\]
where we have used Eqs~(\ref{bsbb}) and (\ref{equip}), which yields
$F^2=\langle\Pgas^2\rangle/\N^2=(\BB^2+\langle\bb^2\rangle)/(4\pi\N^2)$. We
have also introduced the volume filling factor of thermal electrons,
$f=\N^2/\langle\nel^2\rangle=\N^2/(\N^2+\langle\n^2\rangle)$.


\begin{thebibliography}{99}


\bibitem[1997]{B97} Beck, R. 1997, in The Physics of Galactic Halos,
eds H.~Lesch, R.-J.\ Dettmar, U.\ Mebold \& R.\ Schlickeiser
(Akademie Verlag, Berlin), p.~135

\bibitem[2000]{B00} Beck, R. 2000,
Phil.\ Trans.\ R.\ Soc.\ London, Ser.\ A, 358, 777

\bibitem[2001]{B01} Beck, R. 2001, Sp.\ Sci.\ Rev., 99, 243

\bibitem[1996]{BBMSS96} Beck, R., Brandenburg, A., Moss, D., Shukurov,
A., \& Sokoloff, D. 1996, \araa,  34, 155


\bibitem[1999]{B99} Berkhuijsen, E.~M. 1999, in
Plasma Turbulence and Energetic Particles in Astrophysics, eds M.\ Ostrowski
\& R.\ Schlickeiser (Univ.\ Jagiellonski, Krak\'ow), p.~61

\bibitem[1990]{BC90} Boulares, A., \& Cox, D.~P. 1990, \apj, 365, 544


\bibitem[2001]{BT01} Brown, J.~C., \& Taylor, A.~R. 2001, \apj, 563, L31

\bibitem[1966]{b66} Burn, B.~J. 1966, \mnras, 153, 67


\bibitem[1997]{D97} Duncan, A.~R., Haynes, R.~F., Jones, K.~L., \&
Stewart R.~T. 1997, MNRAS, 291, 279

\bibitem[1995]{GS95} Goldreich, P., \& Sridhar, S. 1995, \apj, 438, 763

\bibitem[1997]{GS97} Goldreich, P., \& Sridhar, S. 1997, \apj, 485, 680

\bibitem[1969]{FGH69} Field, G.~B., Goldsmith, D.~W., \& Habing, H.~J. 1969,
\apj, 155, L149

\bibitem[2001]{FS01} Fletcher, A., \& Shukurov, A. 2001, \mnras, 325, 312


\bibitem[2001]{F01} Frick, P., Stepanov, R., Shukurov, A., \& Sokoloff, D.
2001, \mnras, 325, 649

\bibitem[2001]{G01} Gaensler, B.\ M., Dickey, J.\ M.,
McClure-Griffiths, N.~M., Green, A.~J., Wieringa, M.~H., \& Haynes R.~F.
2001, \apj, 549, 959

\bibitem[2001]{GVSS01} Gazol, A., V\'azquez-Semadeni, F., S\'anchez-Salcedo,
F.J. \& Scalo, J. 2001, \apj, 557, L121

\bibitem[2000]{H00} Haverkorn, M., Katgert, P., \& de Bruyn, A.~G.
2000, A\&A, 356, L13

\bibitem[2003]{H03} Haverkorn, M., Katgert, P., \& de Bruyn, A.~G.
2003, A\&A, 403, 1031, and A\&A, 403, 1045

\bibitem[1976]{H76} Heiles, C. 1976, \araa, 14, 1

\bibitem[1996]{H96} Heiles, C. 1996, in Polarimetry of the Interstellar
Medium, eds W.~G.~Roberge \& D.~C.~B.~Whittet, ASP Conf. Ser.~97, 457

\bibitem[2001]{JT01} Jenkins, E.~B., \& Tripp, T.~M. 2001, \apj, 137, 297

\bibitem[1995]{KK95} Kalmykov, N.~N., \& Khristiansen, G.~B. 1995,
J.\ Phys. G: Nucl.\ Part.\ Phys., 21, 1279

\bibitem[1999]{K99} Korpi, M.~J., Brandenburg, A., Shukurov, A., Tuominen,
I., \& Nordlund,  \AA. 1999, \apj,  514, L99

\bibitem[1981]{L81} Laing, R.~A. 1981, \apj, 248, 87

\bibitem[2002]{L02} Laing, R.~A. 2002, \mnras, 329, 417

\bibitem[1970]{L70} Lerche, I. 1970, ApSpSci, 6, 481

\bibitem[1994]{L94} Longair, M.\ S. 1994, High Energy Astrophysics
(Cambridge Univ.\ Press, Cambridge)

\bibitem[1977]{MO77} McKee, C.\ \& Ostriker, J. 1977, \apj,  218, 148

\bibitem[2003]{M2003} Mitra, D.,  Wielebinski, R., Kramer, M., \&
Jessner, A. 2003, A\&A, 398, 993

\bibitem[1993]{OS93} Ohno, H., \& Shibata, S. 1993, \mnras, 262, 953

\bibitem[1979]{P79} Parker, E.~N. 1979, Cosmical Magnetic Fields
(Clarendon Press, Oxford)

\bibitem[1981]{P81} Phillipps, S., Kearsey, S., Osborne, J.~L.,
Haslam, C.~G.~T., \& Stoffel, H. 1981, A\&A, 103, 405

\bibitem[1994]{RL94} Rand, R.~J., \& Lyne, A.~G. 1994, \mnras, 268, 497

\bibitem[1995]{RBK96} Rosen, A., Bergman, J.~N., \& Kelson, D.~D. 1996, \apj, 470, 839

\bibitem[1988]{RSS88} Ruzmaikin, A.~A., Shukurov, A.~M., \& Sokoloff,
D.~D. 1988, Magnetic Fields of Galaxies (Kluwer, Dordrecht).

\bibitem[1982]{S82} Sarkar, S. 1982, MNRAS, 199, 97

\bibitem[1998]{S98} Sokoloff, D.\ D., Bykov, A.\ A., Shukurov, A.\ et
al. 1998, \mnras, 299, 189 (Erratum, \mnras, 303, 207, 1999)

\bibitem[1984]{S84} Spoelstra, T.~A.~T. 1984, A\&A, 135, 239

\bibitem[2000]{S00} Strong, A.~W., Moskalenko, I.~V., \& Reimer, O.
2000, \apj, 537, 763


\bibitem[1999]{U99} Uyan{\i}ker, B., F\"urst, E., Reich, W., Reich, P., \&
Wielebinski, R. 1999, A\&AS, 138, 31



\end{thebibliography}
\end{document}